\documentstyle[12pt]{article}
\def\bea{\begin{eqnarray}}      \def\eea{\end{eqnarray}}
\def\beq{\begin{equation}}      \def\eeq{\end{equation}}
\author{Antonio Candiello\thanks{Supported
in part by M.P.I. This work is carried out in the framework of
the European Community Programme ``Gauge Theories, Applied Supersymmetry
and Quantum Gravity'' with a financial contribution under contract SC1-CT92
-D789.}\\
Dipartimento di Fisica, Universit\`a di Padova\\
Istituto Nazionale di Fisica Nucleare, Sezione di Padova\\
Italy}
\date{}
\title{
\rightline{\protect{{\normalsize DFPD/93/TH/76}}}
\rightline{\protect{{\normalsize hep-th/9401082}}}
\vskip1truecm
{\tt WBase}: a C package to reduce tensor products of Lie algebra
representations}
\begin{document}
\maketitle
\begin{abstract}
It is nearly twenty years that there exist computer programs to reduce
products of Lie algebra irreps. This is a contribution in the field that uses
a modern
computer language (``C'') in a highly structured and object-oriented way.
This gives the benefits of high portability, efficiency, and makes it easier
to include the functions in user programs.
Characteristic of this set of routines is the {\it all-dynamic} approach for
the use of memory, so that the package only uses the memory resources as
needed.
\end{abstract}
\newpage
\subsection*{PROGRAM SUMMARY}
\newdimen\remskip
\newdimen\remindent
\remindent=\parindent
\remskip=\parskip
\parindent0pt
\parskip6pt
{\it Title of program:} {\tt WBase}\par
{\it Catalogue number:}\par
{\it Program obtainable from:} CPC Program library, Queen's University of
Belfast, N.~Ireland (see application form in this issue)\par
{\it Licensing provisions:} none\par
{\it Computer:} Vax 4000/90, DecSystem 5500; {\it Installation:} Department of
Physics, Padua, Italy\par
{\it Operating system:} Vms 5.5-2, Ultrix V4.3A\par
{\it Programming language used:} Vax C without any extension\par
{\it Memory required to execute with typical data:} the program uses dynamic
allocation, the memory used can be just a few words up to the available memory
on the hardware\par
{\it No. of bits in a word:} 32\par
{\it No. of lines in distributed program, including test data, etc:} 875
lines\par
{\it Keywords:} Lie algebras, tensor products, irreducible representations\par
{\it Nature of physical problem}\\
Irreducible representation manipulations are needed in a wide number of
physical arguments, such as high energy particle physics, supergravity
theories, grand unification models\par
{\it Method of solution}\\
The Dynkin technique for manipulating Lie algebra irreps is used. A
sophisticated dynamic allocation scheme let our program to use memory only
when really needed\par
{\it Restrictions of the complexity of the problem}\\
The limit on the application of our program is related to the dimension
of the irreducible representations needed when computing tensor products; this
limit is dependent on the hardware, as dynamic allocation does not set
software bounds to the size of data\par
{\it Typical running time:} 2.41 s for the test run\par
{\it References}\newline
[1] B.~W.~Kernighan and D.~M.~Ritchie, ``The C Language''
(Prentice-Hall, 1978.\par
\parindent\remindent
\parskip\remskip
\section{Introduction}
The necessity of group representation theory in several sectors of physics is
well known in the community. In particular, in particle physics, thanks to the
grand unification models, it is now nearly a necessity to work out some group
representation manipulations. Another sector where representation theory is
heavily used is the construction of supergravity theories; this is in
particular a
sector in which we are involved and the main reason for us for constructing
such a package.

Consulting standard references on the subject \cite{SLANSKY,CAHN,PATERA},
in particular \cite{SLANSKY}, one finds tables of dimensions of irreducible
representations (irreps) and
tables with the reduction of tensor products of irreps. But in certain cases
cases the
tables are not large enough, and then we have two possible choices:
a) do the calculations needed with tools like Young tableaux;
b) search for a computer program that does the job.
Opting for the second choice (a necessity for some Lie groups and for irreps
above certain
dimensions), one can refer to the works of Patera et al \cite{PATERACOMP}.
Our philosophy has been to restart with the C language \cite{KRITCHIE} and,
thanks to the C
powerful dynamic data allocation facilities, to keep all needed data
structures run-time allocated;
we also decided to make use of the most modern object-oriented techniques
in order to free the user
of our package from the complexities of the dynamic allocation scheme, and to
keep the source code clear and easily modifyable (see,
for example, \cite{LIPPMAN,STROUSTRUP}).
The advantages are:
\begin{itemize}
\item dynamic allocation: the routines work on all machines, and memory is
used only when really needed; this is a necessity, as memory requirements can
vary from a few bytes to several megabytes according to the tensor products
needed.
Dynamic allocation gives us also the possibility to restart the routines
changing the algebra without exiting the program, permitting the use of this
package as the engine in an irrep products table generator;
\item object-oriented technology: it is thanks to this approach that we can
keep tractable such a system with continuous allocations and deallocations of
memory, and as an added advantage we can give the user a consistent and easy
interface to our routines.

As an example we cite the clarity of this fragment of C source code:
\begin{verbatim}
pcurr pc; wplace *p;
wblock *base;
for(p=pstart(base,&pc);p!=NULL;p=pnext(&pc))
  wdisp(p->vect);
\end{verbatim}
where the entire irrep weight system, pointed by {\tt base}, is displayed to
the standard output. Note that there is nothing here to know about the
internal data structure of {\tt base}.
\end{itemize}
\vskip 1truecm
The work is organized as follows: in section 2 we describe some of the
intricacies of the Dynkin view of representation theory, where the fundamental
object is the ``weight vector''; section 3 is devoted to the data structures,
their links and allocation/deallocation schemes. In section 4 we enter in the
core of the package, describing the higher level routines which
\begin{itemize}
\item compute the dimension of an irrep
\item compute the casimir index of an irrep
\item generate the weight system of an irrep
\item compute the degeneration of each weight in the weight system
\item gives the reduction of tensor products of two given irreps.
\end{itemize}
Section 5 describes the user-interaction routines and a simple session at the
terminal; we stress that these interaction routines can be easily substituted
by more sophisticated ones, for example batch-oriented routines to generate
full tables, or GUI- (Graphical User Interface) oriented routines to simplify
the task of using the routines to the casual user.
In section 6 we discuss the performance of the package for both the points of
view of 1) speed and 2) memory; they are somewhat connected in that, due to
our linked list scheme, more used memory means more time to process the data.
In this last section we also discuss the direction of future improvements of
the package for both problems of memory and speed.
\section{Dynkin's approach to representation theory}
We will try to keep this section as short as possible, focusing mainly on the
algorithms needed for our purposes; all material is derived from
\cite{SLANSKY}, to which we refer also for a comprehensive introduction to the
subject.

As is well known, a simple Lie algebra is described by his set of generators
which have the commutation rules
\beq
[T_i,T_j]=f_{ijk}T_k,\qquad i,j,k=1,\ldots,d
\eeq
where the $f_{ijk}$ are the algebra structure constants; a convenient basis for
the generators is the Cartan-Weyl basis, where the $T_i$ are subdivided in the
$l$ simultaneously diagonalizable generators $H_i$,
\beq
[H_i,H_j]=0\qquad i=1,\ldots,l
\eeq
and the remaining generators $E_\alpha$,
\beq
[H_i,E_\alpha]=\alpha_iE_\alpha,\qquad i=1,\ldots,l;\
\alpha=-\frac{d-l}{2},\ldots,\frac{d-l}{2}
\eeq
so that the structure constants are now organized as a set of $l$-vectors
$\alpha$: $l$ is the {\it rank} of the algebra, and the set of $l$-vectors
$\alpha$ are the {\it roots}. It results that all roots can be constructed
via linear combinations by a set of $l$ roots, called {\it simple roots}.

With the simple roots one then constructs the $l\times l$ {\it Cartan matrix}
which is needed in our package to construct the weight system of a given irrep,
as we will see; the Cartan matrix is the key to the classification of the Lie
algebras, and it is known for all of them: the $A_n$, $B_n$, $C_n$, $D_n$
series and the exceptional algebras $G_2$, $F_4$, $E_6$, $E_7$, $E_8$. The
other matrix needed is the {\it metric } $G_{ij}$, which is related to the
inverse of the Cartan matrix.

The Dynkin approach is so powerful because the Cartan matrix is all we need to
completely describe the algebra. This is at the basis of
our package: the routine {\tt WStartup}, given the name of
the algebra\footnote{actually only the $A_n$, $B_n$ and $D_n$ series, through
{\tt Afill}, {\tt Bfill} and {\tt Dfill} are implemented, but it is really a
simple task to add support for the other algebras.}, takes care of generating
algorithmically the related Cartan matrix, along with the metric and two
weight vectors we need also.

The metric matrix (which we call {\tt wmetr}) introduces a scalar product in
the space of {\it weight vectors}, which are $l$-uples of integer numbers
describing either irreps or the different states in a given irrep. Each
irrep in an algebra is uniquely classified by such an $l$-uple, the {\it
highest weight} whose components are all positive integers (the Dynkin labels);
both the weight
vectors and the highest weight of an irrep are associated to the typedef {\tt
wvect}.

The dimension of an irrep $\Lambda$ can be calculated with the help of the {\it
Weyl formula}, which is
\beq
\dim(\Lambda)=\prod_{pos. roots \alpha}
\frac{(\Lambda+\delta,\alpha)}{(\delta,\alpha)}
\eeq
where $\Lambda$ is the highest weight determining the irrep, $\delta$ is a
special vector which has Dynkin components which are all equal to unity, and
$(\,,\,)$
is the scalar product constructed with the metric $G_{ij}$. The {\it positive
roots} can be obtained with a specific algorithm which will be explained below
in the context of weight systems: specifically, we derive the weight system of
the adjoint irrep keeping only the positive weight vectors (the upper half of
the spindle). The weyl formula is implemented by the {\tt weyl} function.

Sometimes it is necessary to know other invariants of the irreps, apart from
the dimension; we have implemented the second order casimir invariant
\beq
\hbox{cas}(\Lambda)=\frac{\dim(\Lambda)}{\dim(\hbox{adj})}
\eeq
which is calculated by the function {\tt casimir}.

The {\it weight system} of an irrep is made up by a sequence of weight
vectors, which describe all the states, degenerated or not, in an irrep; the
weight system is also needed to work out products of irreps, which is the main
task of our package. The weight system is obtained by subtracting rows of the
Cartan matrix from the highest weight as described by the following procedure:
\vskip .5truecm
{\em
start with the highest weight: $u:=\Lambda$;\par
\hskip.5truecm{\rm for all $i$ in $1,\ldots,l$}\par
\hskip.9truecm subtract from $u$ the $i$-th row of the Cartan matrix
once, twice, $\ldots,u[i]$\par
\hskip1.4truecm times to give the vectors $u_1,\ldots,u_{u[i]};$\par
\hskip.9truecm add all $u_k$ to the weight system\par
\hskip.9truecm restart the procedure with $u:=u_k$}
\vskip .5truecm
For example, let us see what happens for the {\bf3} of $SU(3)$: given the
Cartan matrix
$\left(\begin{array}{cc}
2&-1\\
-1&2
\end{array}\right)$,
and the highest weight (0 1) of the {\bf3} irrep, we have:
\begin{description}
\item[(0 1)] $\rightarrow$ \fbox{(0 1)}
\begin{itemize}
\item first component $=0$ $\Rightarrow$ no subtractions of (2 -1)
\item second component $=1$ $\Rightarrow$ one subtraction of (-1 2):
 it gives (0 1)$-$(-1 2)$=$\fbox{(1 -1)};
\begin{description}
\item[(1 -1)] $\rightarrow$
\begin{itemize}
\item first component $=1$ $\Rightarrow$ one subtraction of (2 -1):
 it gives (1 -1)-(2 -1)$=$\fbox{(-1 0)}
\begin{description}
\item[(-1 0)] $\rightarrow$ {\em no further subtractions}
\end{description}
\item second component $=-1<0$ $\Rightarrow$ no subtractions of (-1 2).
\end{itemize}
\end{description}
\end{itemize}
\end{description}
So we have the weight system
\beq
\left[\begin{tabular}{cc}
0&1\\
1&-1\\
-1&0
\end{tabular}\right].
\eeq
For higher ranks and/or weights the weight system grows rapidly, but the
steps are essentially the same. The routine that takes care of constructing
the weight system is {\tt wtree}; it is a recursive routine that uses the
procedure described.

The trickier part of the construction of the w.s. results from the fact that
each weight
vector in the w.s. can be degenerated; the degeneration is calculated using
the {\it Freudenthal recursion formula}, level by level, knowing the
degeneration of previous levels: it reads
\beq
\hbox{deg}(w)=\frac{
 2\sum_{pos. roots \alpha}\sum_{k>0}
 \hbox{deg}(w+k\alpha)\,(w+k\alpha,\alpha)
 }{\parallel\Lambda+\delta\parallel^2-\parallel w+\delta\parallel^2}
\eeq
with the initial condition $\hbox{deg}(\Lambda)=1$. This calculation is
performed by the routine {\tt freud} for each weight vector.

Once developed the machinery to generate full weight systems along with their
degenerations, we can reduce the products of irreps, i.e. solve for the $R_i$
in the equation
\beq
R_1\otimes R_2=\bigoplus_i R_i;
\eeq
clearly, as each irrep is individuated by a specific highest weight, this
statement can be re-expressed by saying that, given two highest weight
vectors, we have to find the list of h.w. vectors, each with its degeneration.
The algorithm is simple, but requres many computations; here we present the
main steps along with the related routines called:
\begin{itemize}
\item ({\tt wsyst}): generate the w.s. of $R_1$;
\item ({\tt wsyst}): generate the w.s. of $R_2$;
\item ({\tt bprod}): construct the set $\{w_1+w_2\}$,
$w_1\in\hbox{w.s.}(w_1)$, $w_2\in\hbox{w.s.}(w_2)$;

repeat until the set is empty:
\begin{itemize}
\item ({\tt whighest}): find the h.w. $w$ of the set (it is the weight vector
$w$ which gives the maximum $\overline{R}\cdot w$, where $\overline{R}$ is a
known vector specific to each algebra which we call {\tt whigh}).
\item ({\tt wsyst}): generate the w.s. related to $w$;
\item {(\tt bremove}): subtract from the set each vector in this w.s.
\end{itemize}
\end{itemize}
This is precisely the sequence followed by the routine {\tt wpdisp}.
\section{Dynamic data structures}
The fundamental data structure underlying our routines is the weight vector,
which consists of an $l$-uple of small (small, as the weight vector components
grow slowly with respect to the dimension of the irrep) integer numbers. In a
conservative approach one would use the following definition,
\begin{verbatim}
#define MAX 10
typedef int wvect[MAX];
\end{verbatim}
to introduce the related weight vector data type. However, this definition is
redundant: firstly we can note that an {\tt int} is really more than needed,
it suffices to use the {\tt char} type for the components of the weight
vector. Secondly, the fixed dimension of {\tt wvect} really waste memory when
the rank $l$ is less than {\tt MAX} and, on the other hand, it limits the
maximum rank to {\tt MAX}, so we will need to recompile to use larger rank
algebras.

To overcome these drawbacks, we use the dynamic allocation features of the C
language, defining
\begin{verbatim}
typedef char *wvect;
\end{verbatim}
and introducing the related allocation/deallocation routines
\begin{verbatim}
wvect walloc();
void wfree(wvect w);
\end{verbatim}
which alloc and free a vector of length {\tt wsize}$\equiv l$ which can be
changed at run-time.

For weight vectors embedded in the weight system of an irrep, however, we would
require also some other
information, such as the degeneration and the
level of the weight vector, so we are led to an extended weight structure
which we call {\tt wplace}:
\begin{verbatim}
typedef struct tplace {
 tdeg deg;
 tlevel level;
 char vect[1];} wplace;
typedef short tlevel;
typedef unsigned short tdeg;
\end{verbatim}
where we keep separated the {\tt typedef}s for {\tt deg} and {\tt level} in
order to have easy expandibility (the {\tt short} for {\tt tlevel} works well
for irreps with height less than nearly 32000). The {\tt wplace} data type is
used essentially to give a structure to the raw data kept in the blocks on
which we base our granularity-flexible allocation scheme; also, {\tt wplace}
is the standard object managed by the routines {\tt pstart}/{\tt pnext}/{\tt
plast} which hide the dynamic block allocation scheme.
\subsection{Mid-level routines}
In order to manipulate the lists of blocks that contain the weight systems all
that the user needs is a pointer, say {\tt base}, of type {\tt wblock}, that
contains the weight system of the given irrep. Then it suffices to:
\begin{enumerate}
\item declare an object of type {\tt pcurr} (say, {\tt pcurr pc;})
\item use the iteration functions {\tt pstart}/{\tt pnext} as in
\begin{verbatim}
for(p=pstart(base,&pc);p!=NULL;p=pnext(&pc))
  < do something with p >
\end{verbatim}
remembering that: {\tt p->vect} gives the weight vector, {\tt p->deg} its
degeneration, and {\tt p->level} the level of the vector within the weight
system;
\item to remove the last entry from the list one uses
{\tt base=plast(p,base)}, with the just removed vector returned in the area
pointed by {\tt p};
\item the construction of the w.s. list will be normally handled by the higher
level routines to be described in section 4.
\end{enumerate}
\subsection{Inner data structure}
The difficulty in constructing weight systems of arbitrary irreps is due
to the fact that the dimension of the table needed to store the weight system
is not known in advance. The only solution that does not waste large amounts
of memory and does not limit our routines more than the hardware does is a
multiple block allocation scheme.

The data structure is as follows:
\medskip

% list picture
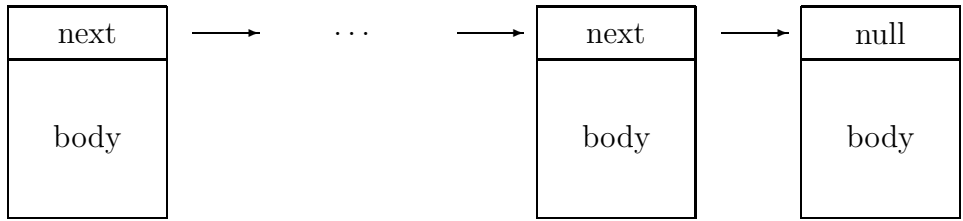
\begin{picture}(500,100)
\put(0,100){\line(1,0){60}}
\put(0,80){\line(1,0){60}}
\put(0,20){\line(1,0){60}}
\put(0,20){\line(0,1){80}}
\put(60,20){\line(0,1){80}}
\put(30,90){\makebox(0,0){next}}
\put(30,50){\makebox(0,0){body}}
\put(70,90){\vector(1,0){25}}

\put(100,90){\makebox(60,0){$\ldots$}}
\put(170,90){\vector(1,0){25}}

\put(200,100){\line(1,0){60}}
\put(200,80){\line(1,0){60}}
\put(200,20){\line(1,0){60}}
\put(200,20){\line(0,1){80}}
\put(260,20){\line(0,1){80}}
\put(230,90){\makebox(0,0){next}}
\put(230,50){\makebox(0,0){body}}
\put(270,90){\vector(1,0){25}}

\put(300,100){\line(1,0){60}}
\put(300,80){\line(1,0){60}}
\put(300,20){\line(1,0){60}}
\put(300,20){\line(0,1){80}}
\put(360,20){\line(0,1){80}}
\put(330,90){\makebox(0,0){null}}
\put(330,50){\makebox(0,0){body}}
\end{picture}

a singly linked list of blocks of identical size fixed at run-time (according
to the fragmentation required) by the number of {\tt wplace} vector entries as
given in {\tt bsize}; the typedef that defines a single block leaves therefore
its main structure undefined, as it is different for different ranks {\tt
wsize} and block dimension {\tt bsize}:
\begin{verbatim}
typedef struct tblock {
 struct tblock *next;
 char body[1];} wblock;
\end{verbatim}
also this definition, as the {\tt wplace} one, is used through casts that give
form to unstructured raw data as returned by the allocator {\tt balloc}.
Single blocks are freed with a call to {\tt bfree}, linked blocks are freed
with a call to {\tt bsfree} acted on the first of them.
The structured form of the blocks, when casting with {\tt wplace} and defining
{\tt bsize} and {\tt wsize}, is as follows:
\medskip

% list picture with vects & degs
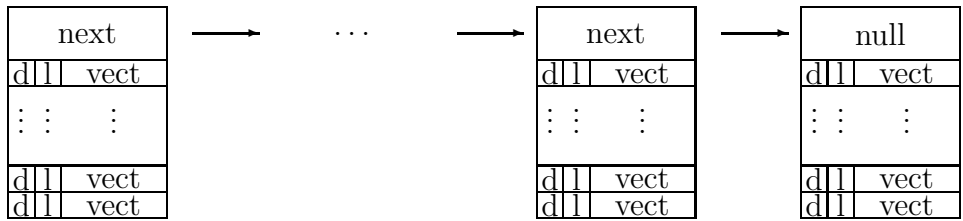
\begin{picture}(500,100)
\put(0,100){\line(1,0){60}}
\put(0,80){\line(1,0){60}}
\put(0,20){\line(1,0){60}}
\put(0,20){\line(0,1){80}}
\put(60,20){\line(0,1){80}}
\put(30,90){\makebox(0,0){next}}
\put(0,70){\line(1,0){60}}
\put(0,40){\line(1,0){60}}
\put(0,30){\line(1,0){60}}
\put(10,70){\line(0,1){10}}
\put(20,70){\line(0,1){10}}
\put(10,30){\line(0,1){10}}
\put(20,30){\line(0,1){10}}
\put(10,20){\line(0,1){10}}
\put(20,20){\line(0,1){10}}
\put(5,75){\makebox(0,0){d}}
\put(15,75){\makebox(0,0){l}}
\put(40,75){\makebox(0,0){vect}}
\put(5,35){\makebox(0,0){d}}
\put(15,35){\makebox(0,0){l}}
\put(40,35){\makebox(0,0){vect}}
\put(5,25){\makebox(0,0){d}}
\put(15,25){\makebox(0,0){l}}
\put(40,25){\makebox(0,0){vect}}
\put(5,60){\makebox(0,0){$\vdots$}}
\put(15,60){\makebox(0,0){$\vdots$}}
\put(40,60){\makebox(0,0){$\vdots$}}
\put(70,90){\vector(1,0){25}}

\put(100,90){\makebox(60,0){$\ldots$}}
\put(170,90){\vector(1,0){25}}

\put(200,100){\line(1,0){60}}
\put(200,80){\line(1,0){60}}
\put(200,20){\line(1,0){60}}
\put(200,20){\line(0,1){80}}
\put(260,20){\line(0,1){80}}
\put(230,90){\makebox(0,0){next}}
\put(200,70){\line(1,0){60}}
\put(200,40){\line(1,0){60}}
\put(200,30){\line(1,0){60}}
\put(210,70){\line(0,1){10}}
\put(220,70){\line(0,1){10}}
\put(210,30){\line(0,1){10}}
\put(220,30){\line(0,1){10}}
\put(210,20){\line(0,1){10}}
\put(220,20){\line(0,1){10}}
\put(205,75){\makebox(0,0){d}}
\put(215,75){\makebox(0,0){l}}
\put(240,75){\makebox(0,0){vect}}
\put(205,35){\makebox(0,0){d}}
\put(215,35){\makebox(0,0){l}}
\put(240,35){\makebox(0,0){vect}}
\put(205,25){\makebox(0,0){d}}
\put(215,25){\makebox(0,0){l}}
\put(240,25){\makebox(0,0){vect}}
\put(205,60){\makebox(0,0){$\vdots$}}
\put(215,60){\makebox(0,0){$\vdots$}}
\put(240,60){\makebox(0,0){$\vdots$}}
\put(270,90){\vector(1,0){25}}

\put(300,100){\line(1,0){60}}
\put(300,80){\line(1,0){60}}
\put(300,20){\line(1,0){60}}
\put(300,20){\line(0,1){80}}
\put(360,20){\line(0,1){80}}
\put(330,90){\makebox(0,0){null}}
\put(300,70){\line(1,0){60}}
\put(300,40){\line(1,0){60}}
\put(300,30){\line(1,0){60}}
\put(310,70){\line(0,1){10}}
\put(320,70){\line(0,1){10}}
\put(310,30){\line(0,1){10}}
\put(320,30){\line(0,1){10}}
\put(310,20){\line(0,1){10}}
\put(320,20){\line(0,1){10}}
\put(305,75){\makebox(0,0){d}}
\put(315,75){\makebox(0,0){l}}
\put(340,75){\makebox(0,0){vect}}
\put(305,35){\makebox(0,0){d}}
\put(315,35){\makebox(0,0){l}}
\put(340,35){\makebox(0,0){vect}}
\put(305,25){\makebox(0,0){d}}
\put(315,25){\makebox(0,0){l}}
\put(340,25){\makebox(0,0){vect}}
\put(305,60){\makebox(0,0){$\vdots$}}
\put(315,60){\makebox(0,0){$\vdots$}}
\put(340,60){\makebox(0,0){$\vdots$}}
\end{picture}

\section{High level routines}
Given the underlying data structure, the task of handling the lists of vectors
is taken by a few routines:
\begin{itemize}
\item {\tt wsave(w,base,level)}: \par
insert in the list pointed by {\tt base} the weight vector {\tt w} before any
weight vector at the same or higher level, if {\tt w} is not already present;
\item {\tt wremove(w,base)}:\par
remove a weight vector {\tt w} from the list {\tt base}, counting the
degeneration, i.e. by decrementing the degeneration associated, if greater
than 1, and removing the vector only if the degeneration is just 1;
\item {\tt wtree(w,base,level)}:\par
construct, repeatedly calling {\tt wsave} and recursively itself, the weight
system vectors of highest weight {\tt w} with the algorithm described in
section 2, without computing the degeneration;
\item {\tt bfreud(base)}:\par
calculate the degeneration for each element in the list through the
Freudenthal formula coded in the function {\tt freud} called for each element
of the weight system in {\tt base}.
\end{itemize}
These last two functions are called in turn by {\tt wsyst(hw)}, which return
the full linked list of blocks of the weight system of highest weight {\tt hw}
along with the degenerations.

Beyond {\tt wsyst}, there is a set of routines to manipulate irreps:
\begin{itemize}
\item {\tt weyl(hw)}, {\tt wdim(base)}\par
both return the dimension of a given irrep, but {\tt weyl} uses the Weyl
formula, while {\tt wdim} requires the construction of the w.s. via {\tt
wsyst} ({\tt wdim} is really a test routine);
\item {\tt casimir(hw)}, {\tt wind(base)}\par
both return the index of a given irrep, but {\tt casimir} uses a formula which
involves only the highest weight {\tt hw} and the positive roots, like the
Weyl formula, while {\tt wind} requires the construction of the w.s. via {\tt
wsyst} (also {\tt wind} is really a test routine);
\item {\tt wheight(hw)}, {\tt whigher(base)}\par
{\tt wheight} returns the height of the irrep of highest weight {\tt hw},
while {\tt whigher} returns the higher weight vector in {\tt base}.
\end{itemize}
The functions {\tt weyl}, {\tt casimir}, {\tt wheight} are called by {\tt
wfdisp(hw)} to give the information about the irrep with h.w. {\tt hw}.
\begin{itemize}
\item {\tt wpdisp(hw1,hw2,mod)}\par
this function hides all the complexities of reducing products of irreps and of
the underlying data structure, by giving to the standard output all the irreps
in the product, from the highest to the lowest, according to the modality
choosen by {\tt mod}\footnote{See the header file {\tt wbase.h}.}. The product
routines are also available as iteration functions {\tt wpstart(b)}/{\tt
wpnext(b,base)}, similar to {\tt pstart}/{\tt pnext}, giving the irreps
contained in the product of each iteration as {\tt wblock} data types.
\end{itemize}
\section{User interaction}
The convenience of the object-oriented approach is visible in the simplicity
of the setting up of a one-page file which contains all that is necessary for
connecting our routines with the user.

This file is {\tt wmain.c} and can be separately compiled; a more sophisticated
user interaction interface may be constructed taking this file as an example.
Let us describe the fundamental routines that need to be called to build up
such a system:
\begin{itemize}
\item {\tt wstartup(type,rank)}, {\tt wcleanup(string)}\par
these are the open/close routines for setting up the algebra; given e.g. the
algebra B5 (SO(11)) one would call {\tt wstartup('B',5)} to allocate the
related vectors and matrices, and then {\tt wcleanup()} to deallocate them.
Each invocation of {\tt wstartup()} must be followed by {\tt wcleanup()}.
Naturally, the
construction of a table of several algebras would call for multiple
invocations of this pair of routines.
\item {\tt wread(hw)}, {\tt wfdisp(hw)}:\par
these are the input/output routines, which read a highest weight vector from
the standard input and display the related irrep information on the standard
output. One has to remember that these routines need a valid {\tt wvect} which
is handled by {\tt walloc()}/{\tt wfree()}.
\item {\tt wsyst(hw)}, {\tt bdisp(base)}, {\tt bsfree(base)}:\par
these routines manage the weight systems:\par
{\tt base=wsyst(hw)} generates the w.s. of {\tt hw} and returns the related
{\tt wblock} pointer; {\tt bdisp(base)} displays to the standard output all
entries of the w.s., and {\tt bsfree(base)} deallocates the complete linked
list of pointers;
\item {\tt wpdisp(hw1,hw2,mod)}:\par
this single routine manages all the intricacies connected to the reduction of
products of irreps, displaying to the standard output all the representations
contained in ``${\tt hw1}\otimes{\tt hw2}$'', according to the {\tt mode}
switch: if the {\tt WTREE} flag is switched on, the full tree of
the found irreps it si showed to the standard output, if the {\tt WFAST} flag
is switched on a faster
and imprecise method of generating products of irreps is used which does not
give correctly the degenerations.
\end{itemize}
The more sophisticated user of our package would need the access to the
iteration functions {\tt wpstart}/{\tt wpnext} which make the skeleton of {\tt
wpdisp}; essentially, {\tt wpdisp} is structured as follows:
\begin{verbatim}
buf=bprod(wsyst(hw1),wsyst(hw2));
for(b=wpstart(buf);b!=NULL;b=wpnext(b,buf))
  ;
\end{verbatim}
where {\tt bprod} takes care of constructing a combined w.s. made from sums of
those of its arguments. The reduction process is hidden by the iteration
routines {\tt wpstart}/{\tt wpnext}, which essentially at each step find the
higher weight in {\tt buf}, generate his w.s., display some information on it,
and remove it from {\tt buf}.
\vskip.8truecm
Let us now turn on the use of our package as it is; it is very simple: after
running the program (wbase) you will be requested for the algebra.
After entering, for example, D5$\equiv$SO(10), a little menu with the possible
operations
will be presented ('P' for product, 'R' for informations on irreps,...).
Selecting 'P', the user will be prompted for two couples of 5 integer numbers
each (the
h.w. of the irreps to be multiplied). With the following input (corresponding
to the {\bf10} and {\bf16} irreps of D5, respectively),
\begin{verbatim}
1 0 0 0 0
0 0 0 0 1
\end{verbatim}
we would have the output
\begin{verbatim}
(D=144,C=68,H=18)  HW=(1 0 0 0 0)
(D=16,C=4,H=10)    HW=(0 0 0 0 1)
\end{verbatim}
which means $10\otimes16=144\oplus\overline{16}$. To associate h.w. to
dimensions, casimirs and levels of the irreps it is useful to choose 'R' as
option. On VMS systems ctrl-z returns the program to the main menu; on the
others, EOF will force the exit.
\section{Performance, memory requirements and future developments}
The package is now, we think, useful for the standard irrep-related
necessities of physicists, as the routines are reasonably fast and the memory
requirements kept to a minimum such as to permit working also with small
computers. We tested the package on both intel- and motorola-powered machines
to prove the usefulness of the package on the personal computer platforms. The
package works better, as it is obvious, when there is more processing power
and memory; another advantage of mainframe-sized machines lies in their batch
execution facilities, useful for greater dimension irreps.

We think that the package can be used by the researchers in both experimental
and theoretical physics as a tool to manage the irrep invariants and
the products of irreps: for that use, with the processing of irreps of
dimension less than 1000, the program gives answers nearly instantaneously,
and the memory requirements are limited; this because the products take a time
proportional to the highest dimension of the reduced irreps. As far as higher
dimensional irreps products are needed, the time required grows rapidly
after dimension 10000 irreps are reached.

Our necessities in supergravity theory required some products of 120 irreps of
D5; so we tested the program in the worst case of the highest irrep in
$120\otimes120$, the 4125, multiplied by another 120. First we obtained (on a
Vax 4000, in batch)
\beq
120\otimes120=4125\oplus5940\oplus1050\oplus\overline{1050}\oplus945
\oplus2\cdot210\oplus54\oplus45\oplus1
\eeq
in nearly 10 minutes of cpu time; then we did the products of each irrep with
another 120:
\bea
\lefteqn{120\otimes4125=70070\oplus192192\oplus48114\oplus34398\oplus48114
\oplus27720}\nonumber\\
&&\oplus36750\oplus2\cdot10560\oplus4312\oplus3696\oplus\overline{3696}
\oplus2970\oplus1728\oplus120
\eea
in nearly 5 hours of cpu time.

This really displays the actual limit of our package: it will process irreps
as long as the product does not exceed about dimension 1000000 (on our
computing facilities), not only because of the time (which grows rapidly) but
because of the memory requested to hold such huge weight systems in memory.

While we think that $10000 \times 10000$ dimension irrep products are really
more than needed by the physics researchers, we report here some of the
improvements we have in schedule to speed up the package and to reduce memory
requirements.
\begin{itemize}
\item First the memory problem: here we have no great cure for this, due to the
simple fact that {\sl a dimension 1000000 irrep really means a
weight system of nearly 100000 vectors}. Recalling that we have already
reduced as possible the
memory required by a {\tt wplace} structure to: \par
$l$ bytes for the weight vector itself,\par
1 short for degeneration,\par
1 short for level (which must be made a long
to permit irreps with level greater than 65000)\par
which gives us the {\it minimum} of $l+4$ bytes per vector; hoping for as much
degeneration as to reduce the need for weight vectors from 1000000 to 100000,
we would have to find, for a 5-rank algebra, one megabyte of memory for a
dimension 1000000 weight system. We really can not
reduce the memory occupied by weight systems without introducing some sort of
in-memory data compression, which will bring us, however, a speed penalty.
\item Regarding the speed constraints, we have more directions, some of
which are already in progress \cite{INPREP}:
\begin{enumerate}
\item Use of the spindle shape property of weight systems. This optimization
relies on the fact that the most part of time taken to construct weight systems
is used in the lower half of the w.s., because of the recursive nature of much
of the routines involved, such as {\tt wtree} and (mostly) {\tt freud} (this
one is the principal cause of slowness, because it calls itself recursively
from higher to lower levels.

The key point lies in the fact that we can truncate the elaboration at half
the height of an irrep (which is known by {\tt wheight}) and generate
algorithmically the second half through a mirroring procedure which is
specific to each algebra.
\item Creation of a {\sl pointer cache} to facilitate the traversing of the
w.s. list by routines such as {\tt freud}; this is a simple and effective
improvement, that will accelerate considerably the w.s. scanning routines. One
has to enlarge the block definition in order to hold a set of \#height
pointers, each of which will point to the first {\tt wplace} element at a each
height.
\item A ``technologic'' improvement can be obtained by rewriting the routines
that actually use {\tt double} arguments into integer routines (the need for
floating point arithmetics derives from the metric matrix, which is related to
the inverse of the Cartan matrix).
\item A user-friendly improvement, useful for batch processing, would be a set
of routines for managing lists of weight systems, such as to facilitate the
construction of multiple products and products of sums of irreps; it will
permit to treat sums of irreps as single entities, and we could directly
compute products such as $(3\oplus6)\otimes(1\oplus8)$.
\end{enumerate}
\end{itemize}
\vskip0.5truecm
\section*{Acknowledgements}
Thanks go to K. Lechner for his training on Lie algebra irrep techniques.

\end{document}